%% file: main.tex
\documentclass{article}
\usepackage{spconf,amsmath,graphicx}

\usepackage{amssymb}
\usepackage{amsfonts}
\usepackage{bm}
\usepackage{todonotes}
\usepackage{subcaption}
\input{pre-packages.tex}
\input{pre-definitions.tex}
\input{pre-declarations.tex}

\title{Tracing Network Evolution using the PARAFAC2 model}
\name{Marie Roald$^1$, Suchita Bhinge$^2$, Chunying Jia$^2$, Vince Calhoun$^3$, T\"{u}lay Adal\i$^2$, Evrim Acar$^1$}
\address{$^{1}$ Simula Metropolitan Center for Digital Engineering, Oslo, Norway \\
$^{2}$ Dept of Computer Science and Electrical Engineering, UMBC, Baltimore, MD \\
$^{3}$ Dept of Psychology, Georgia State University, Atlanta, GA}
\usepackage{varioref}                       
\usepackage[hyperfootnotes=true,hypertexnames=false]{hyperref} 
\usepackage[nameinlink, capitalize, noabbrev]{cleveref} 

\usepackage{booktabs}
\usepackage{makecell}
\usepackage{arydshln}

\newcommand{\diag}{\text{diag}}

\renewcommand{\Mn}[2]{\M{#1}_{#2}} 
\renewcommand{\MnTra}[2]{\M{#1}_{#2}^{\sf T}} 

\begin{document}
\topmargin=0mm
\setlength{\parskip}{0em}
\setlength{\tabcolsep}{5pt}
%
\maketitle
\begin{abstract}
Characterizing time-evolving networks is a challenging task, but it is crucial for understanding the dynamic behavior of complex systems such as the brain. For instance, how spatial networks of functional connectivity in the brain evolve during a task is not well-understood. A traditional approach in neuroimaging data analysis is to make simplifications through the assumption of static spatial networks. In this paper, without assuming static networks in time and/or space, we arrange the temporal data as a higher-order tensor and use a tensor factorization model called PARAFAC2 to capture underlying patterns (spatial networks) in time-evolving data and their evolution. Numerical experiments on simulated data demonstrate that PARAFAC2 can successfully reveal the underlying networks and their dynamics. We also show the promising performance of the model in terms of tracing the evolution of task-related functional connectivity in the brain through the analysis of functional magnetic resonance imaging data.
\end{abstract}
\section{Introduction}
\label{sec:intro}
Time-evolving data analysis is of interest in many disciplines to capture the underlying patterns as well as the evolution of those patterns. For instance, in neuroscience, underlying patterns may correspond to spatial networks capturing functional connectivity \cite{YuLiZh18}, and in social networks, patterns may reveal communities. Capturing those patterns (networks) as well as their temporal evolution, holds the promise to improve our understanding of complex dynamic systems such as the brain, social networks, and molecular mechanisms in the body \cite{WoPaKo18}.

An effective way of representing time-evolving data is to use higher-order tensors (also referred to as multi-way arrays), i.e., tensors with more than two axes of variation. For instance, functional magnetic resonance imaging (fMRI) data from multiple subjects can be rearranged as a third-order tensor with modes: \emph{subjects}, \emph{voxels}, and \emph{time windows}. Tensor factorizations \cite{AcYe09,tensordec:KoTaBa09} have proved useful in terms of revealing the underlying patterns in such higher-order data sets. Previously, for the analysis of time-evolving data, a popular tensor factorization model called the CANDECOMP/PARAFAC (CP) \cite{PARAFAC:Ha70, PARAFAC:Ca70} model has been used to extract temporal patterns to address the temporal link prediction problem \cite{tensor:linkpred:AcDuKo09}, to capture the evolving popularity of different meaningful topics from email threads \cite{tensor:CP:enron:BaBeBr08} and to detect suspicious activity in network traffic \cite{tensor:poisson:ensign:BrBaEzHeLe16}. However, the CP model assumes that underlying patterns stay the same across time, which may not be satisfied by dynamic data \cite{temporal:ShiftCP:MoHaArLiMa08,temporal:ConvCP:MoHa11,tensor:smoothness:parafac2:COPA:AfPePaSeHoSe18,parafac2:BrAnKi99}.

In data mining, there is an increasing interest in capturing time-evolving factors. One approach is to use sliding window-based methods \cite{tensor:slidingwindow:SuPaYu08, tensor:tenClustS:FeFaGa18}; however, patterns are still assumed to be static within each window, and determining the window size is a challenge. 
Alternatively, dynamic data has been modeled using temporal matrix factorizations with temporally evolving patterns \cite{Matrix:community:temporalnetworks:TMF:YuAgWa17, matrix:collaborative:temporalsvd:Ko09}, but with a focus on reconstruction of data matrices, e.g., for link prediction, without discussing the uniqueness of the captured patterns.

In this paper, we use the PARAFAC2 tensor factorization model, with rather well-studied uniqueness properties \cite{KiTeBr99}, to model time-evolving data in such a way that coupled matrices correspond to matrices at different time points and underlying patterns in the matrices may change over time. PARAFAC2 has been used to analyze time-evolving data previously \cite{evolvingfactors:SCAP:TiKi03, parafac2:fMRI:MaChNaMo17}, but the coupled matrices corresponded to matrices for different subjects and patterns were allowed to change across subjects, rather than across time. Here, we assess the performance of the PARAFAC2 model in terms of capturing temporal evolution of the underlying patterns and demonstrate that it is a promising tool to trace the evolution of networks such as growing, shrinking and shifting networks. We also use the PARAFAC2 model in a novel neuroimaging application to understand task-related connectivity and capture time-evolving connectivity patterns that significantly differ between patients with schizophrenia and healthy controls. 

\section{Methods}
\emph{Higher-order tensors} are extensions of matrices to more than two modes, i.e., a vector is a first-order tensor and a matrix is a second-order tensor. This section briefly describes two tensor factorization models, CP and PARAFAC2, which we compare in terms of capturing evolving patterns.\\

{\noindent \bf{CANDECOMP/PARAFAC:}}
The CP model \cite{PARAFAC:Ha70,PARAFAC:Ca70} represents a tensor as the sum of minimum number of rank-one components, i.e., an $R$-component CP model of tensor $\T{X} \in \Real^{I \times J \times K}$ is defined as 
$\T{X} \approx \sum_{r=1}^R \MC{A}{r} \Oprod \MC{B}{r} \Oprod \MC{C}{r}$, where $\Oprod$ denotes the vector outer product, $\MC{A}{r}, \MC{B}{r}$,  and $\MC{C}{r}$ are the $r$th column of factor matrices $\M{A} \in \Real^{I \times R}, \M{B} \in \Real^{J \times R}$, and $\M{C} \in \Real^{K \times R}$, respectively. Using CP, each frontal slice ($\M{X}_k \in \Real^{I \times J}$) is formulated as:
\begin{equation}
\small
    \M{X}_k \approx \M{A} \diag(\MR{c}{k}) \M{B}\Tra,
\end{equation}
where $\diag(\MR{c}{k})$ is an $R \times R$ diagonal matrix with the $kth$ row of $\M{C}$ on the diagonal. If $\T{X}$ represents a \emph{subjects} by \emph{voxels} by \emph{time windows} tensor, each column of $\M{B}$ may reveal a spatial network, $\M{A}$ indicates which networks are present in each subject, and $\M{C}$ contains the temporal profile of each network.\\

\begin{figure}[t]
\begin{minipage}[b]{1.0\linewidth}
  \centering
  \centerline{\includegraphics[width=1.0\linewidth, trim=5 5 10 5,clip]{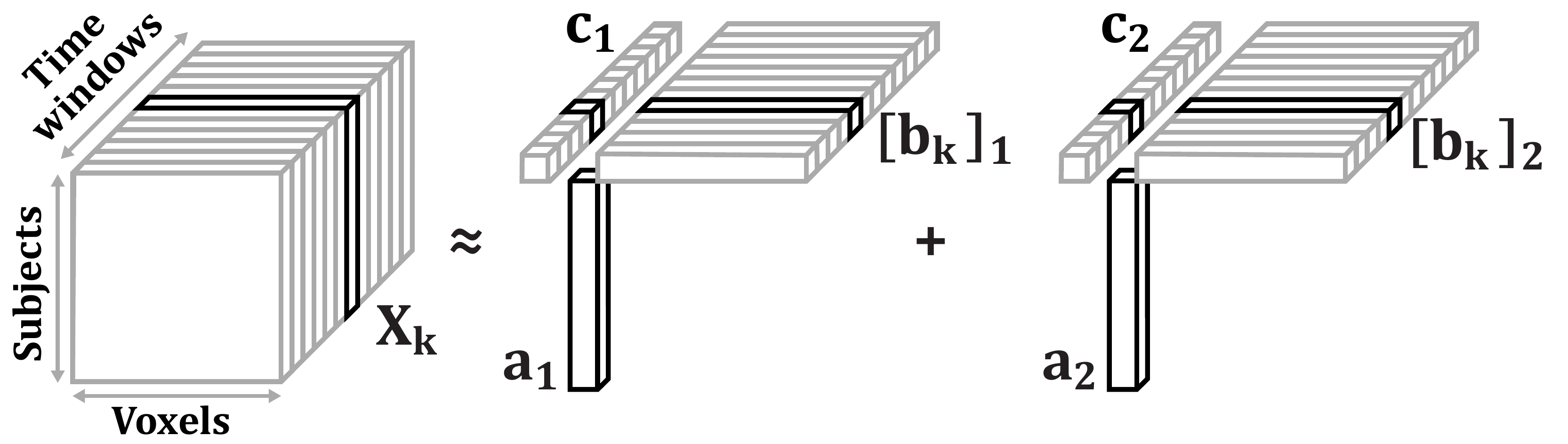}}
\end{minipage}
\caption{\ninept{Illustration of a two-component PARAFAC2 model.}}
\label{fig:multilinearity_parafac2}
\end{figure}

{\noindent \bf{PARAFAC2:}}
PARAFAC2 \cite{tensor:parafac2:Ha72} is a more flexible model than the CP model. While CP assumes that each slice, \(\TFS{X}{k} \in \Real^{I \times J}\), has the same $\M{A}$ and $\M{B}$ matrices, PARAFAC2 allows each slice to have a different $\M{B}$ matrix (\cref{fig:multilinearity_parafac2}) as follows: 
\begin{equation}
\small
    \TFS{X}{k} \approx \M{A} \diag(\MR{c}{k}) \MnTra{B}{k}, \label{eq:pf2}
\end{equation}
where $\Mn{B}{k}$s follow the \emph{PARAFAC2 constraint}, i.e., $\smash{\MnTra{B}{k_1}\Mn{B}{k_1}} = {\MnTra{B}{k_2}\Mn{B}{k_2}}$ for all $k_1, k_2 \leq K$. If $\TFS{X}{k}$ slices correspond to time windows, then networks captured by columns of $\Mn{B}{k}$ may change over time as long as they satisfy this constraint. We use the notation $[\MC{b}{k}]_r$ to denote the $r$th column of matrix $\Mn{B}{k}$.

\section{Experiments}
In this section, we assess the performance of PARAFAC2 in terms of extracting time-evolving networks using both simulated and real data. First, we generate data with time-evolving patterns, and demonstrate that the model can recover the underlying evolving patterns, performing much better than a CP model. Then we use the PARAFAC2 model to analyze fMRI signals from a group of subjects consisting of healthy controls and patients with schizophrenia and show the promise of the model in terms of revealing the difference in the evolution of task-related spatial networks across the two groups.

\subsection{Simulated data}

\subsubsection{Generation of simulated data}\label{sec:simulation}
We generated two types of simulated data: (i) data sets following \eqref{eq:pf2} and the PARAFAC2 constraint, and (ii) data sets following \eqref{eq:pf2}, having time-evolving patterns $\Mn{B}{k}$s that do not necessarily follow the PARAFAC2 constraints.

Factor matrices in each mode are generated as follows (with $R=4$): The factor matrix for the first mode (i.e. subjects) (\(\M{A}\)) has a clustering structure as shown in \cref{fig:AandC}. For the second mode (i.e. voxels), (\(\Mn{B}{k}\))s are generated in two different ways: (i) \emph{Random:} random matrices following the PARAFAC2 constraint, (ii) \emph{Network:} matrices with each column corresponding to either a shifting, a growing, a shrinking or a both shifting and growing network (\cref{fig:B}). Finally, the factor matrix in the third mode (i.e., time) (\(\M{C}\)) is generated in two different ways: (i) \emph{Random:} with all columns drawn from a uniform random distribution, (ii) \emph{Trends:} with one random column, and other three columns following either a sinusoidal, an exponential or a sigmoidal curve (\cref{fig:AandC}).\footnote{Simulations are described in detail in the supplementary material: \url{https://github.com/marieroald/ICASSP20}, with links to the simulation code and Python implementations of CP and PARAFAC2.} 

Once $\T{X}$ is constructed based on \eqref{eq:pf2} using the generated factor matrices, we add random noise: $\T{X}_{\text{noisy}} = \T{X} + \eta\T{E}\frac{\fnorm{\T{X}}}{\fnorm{\T{E}}}$, where $\eta$ is the noise level, $\fnorm{\cdot}$ is the Frobenius norm, entries of $\T{E}$ follow a standard normal distribution. 

\subsubsection{Performance evaluation}
We generated twenty random data sets for all possible combinations of factor matrix generation schemes and used different noise levels; $\eta=0$ and $\eta=0.33$ ($\approx 10\%$ of the data being noise). We fit both CP and PARAFAC2 using alternating least squares \cite{tensordec:KoTaBa09,KiTeBr99} with multiple random starts, and the start with the best fit score was chosen for further analysis (after validating the uniqueness of the model). Non-negativity constraints were imposed in the \emph{time} mode when fitting PARAFAC2 to resolve the sign ambiguity \cite{tensor:sign:parafac2:BrLeJo13}, which is more critical for PARAFAC2 than CP. Model performance is assessed in terms of the following measures:

{\noindent \bf{Model fit:}} One metric used to assess the quality of the data reconstruction, denoted by $\That{X}$, is the \emph{model fit} defined as $ \text{Fit} (\T{X},\That{X}) = 100 \times \left(1 - \frac{ \norm{\T{X} - \That{X}}^2}{ \norm{\T{X}}^2}\right)$. The fit tells us how well the model explains the data. 

{\noindent \bf{Factor Match Score (FMS):}}
We measure the accuracy of the methods, i.e., how well the methods recover underlying factor vectors, using FMS defined, for each mode separately, as:
\begin{equation*}
\small
    \text{FMS}_{\M{U}} = \frac{1}{R}\sum_{r=1} \frac{\left|\MC{u}{r}\Tra \MhatC{u}{r}\right|}{\norm{\MC{u}{r}} \norm{\MhatC{u}{r}}},
\end{equation*}
where $\V{u}_r$ and $\Vhat{u}_r$ are true and estimated $r$th column of factor matrix $\M{U}$, respectively (after fixing the permutation ambiguity). For the evolving mode ($\M{B}$), we concatenate all time steps to form a $(JK\times R)$ matrix, $\tilde{\M{B}}$, and compute $\text{FMS}_{\M{B}}$ using $\tilde{\M{B}}$. For CP, the same $\V{b}_r$ vector is repeated $K$ times.\\
{\noindent \bf{Clustering accuracy:}}
The clustering performance is assessed using $k$-means clustering on the factor matrix extracted from the first mode, using all possible combinations of factor vectors, and the best performance is reported. 

\subsubsection{Results of simulated data analysis}
\cref{tab:results} shows the average performance (for twenty random data sets) of CP and PARAFAC2 ($R=4$) using different data generation schemes. Results demonstrate that PARAFAC2 performs well in terms of recovering the true patterns and their evolution with average FMS values above 0.90 for the evolving network mode, and with much higher FMS in other modes, consistently performing better than the CP model. 

We observe that CP partially discovers $\M{A}$ and $\M{C}$ matrices for the network data, even in the noisy case, while completely failing for data with random $\M{B}$s. Despite failing to capture the true $\M{A}$, CP often has high clustering accuracy due to the clustering structure in the simulations designed to separate groups with several components, i.e., if a model recovers any informative components, the clustering accuracy will be high. 

\begin{figure}[t!]

\begin{minipage}[t]{1.0\linewidth}
  \centering
  \centerline{\includegraphics[width=1.0\linewidth,trim={0 0 0 0.8cm},clip]{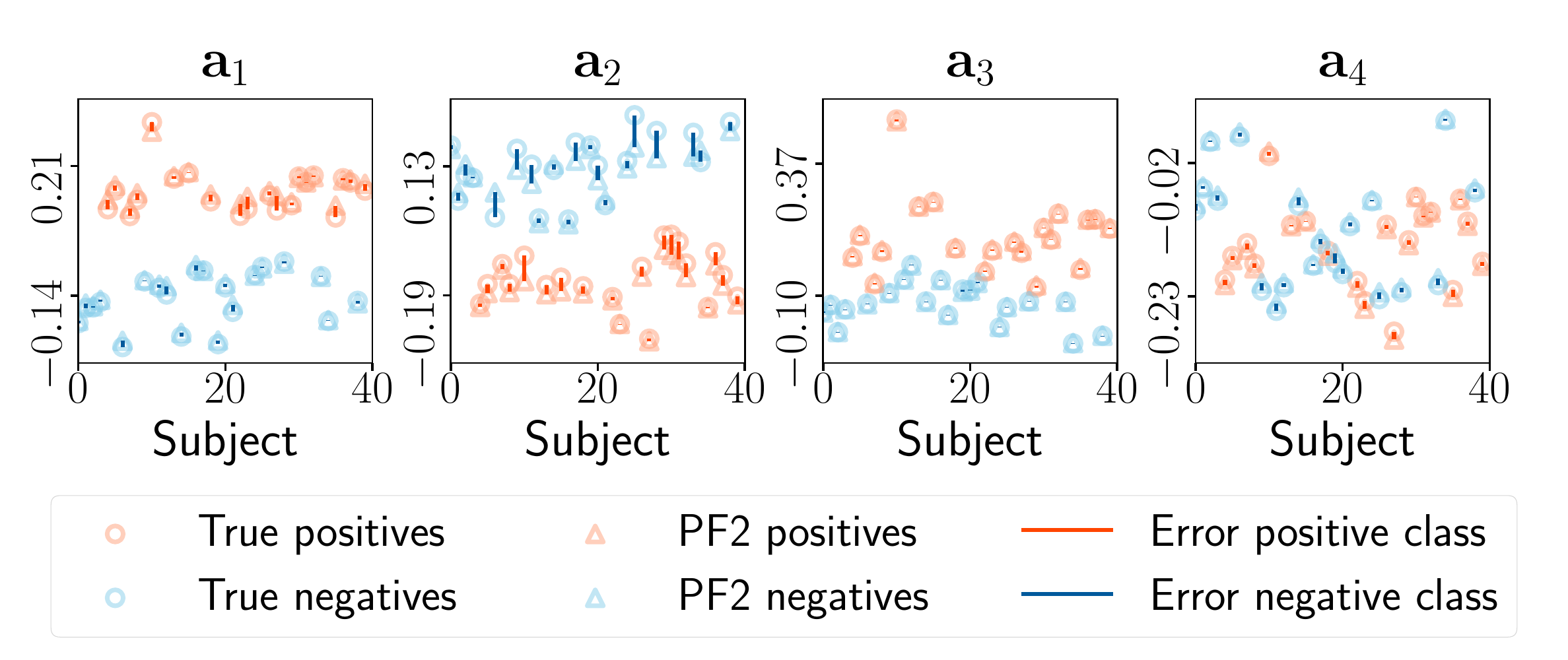}}
  \centering
  \centerline{\includegraphics[width=1.0\linewidth,trim={0 1.2cm 0 0.8cm},clip]{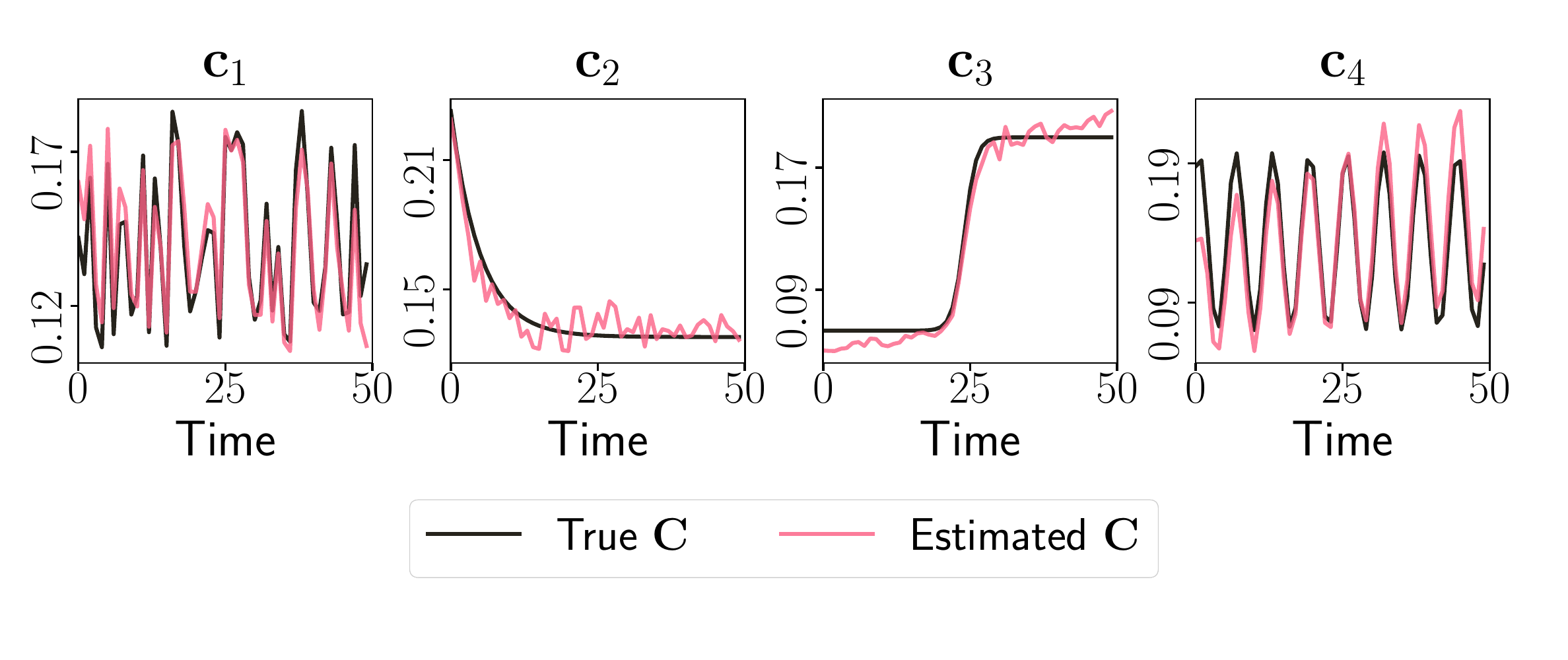}}
\end{minipage}
\caption{\ninept{Top: Scatter plot of true columns of \(\M{A}\) and the ones captured by PARAFAC2. Bottom: Line plot of true columns of \(\M{C}\) and the ones estimated by PARAFAC2 (noisy case).}}
\label{fig:AandC}
\end{figure}

\begin{figure}[t!]

\begin{minipage}[b]{1.0\linewidth}
  \centering
  \centerline{\includegraphics[width=0.95\linewidth]{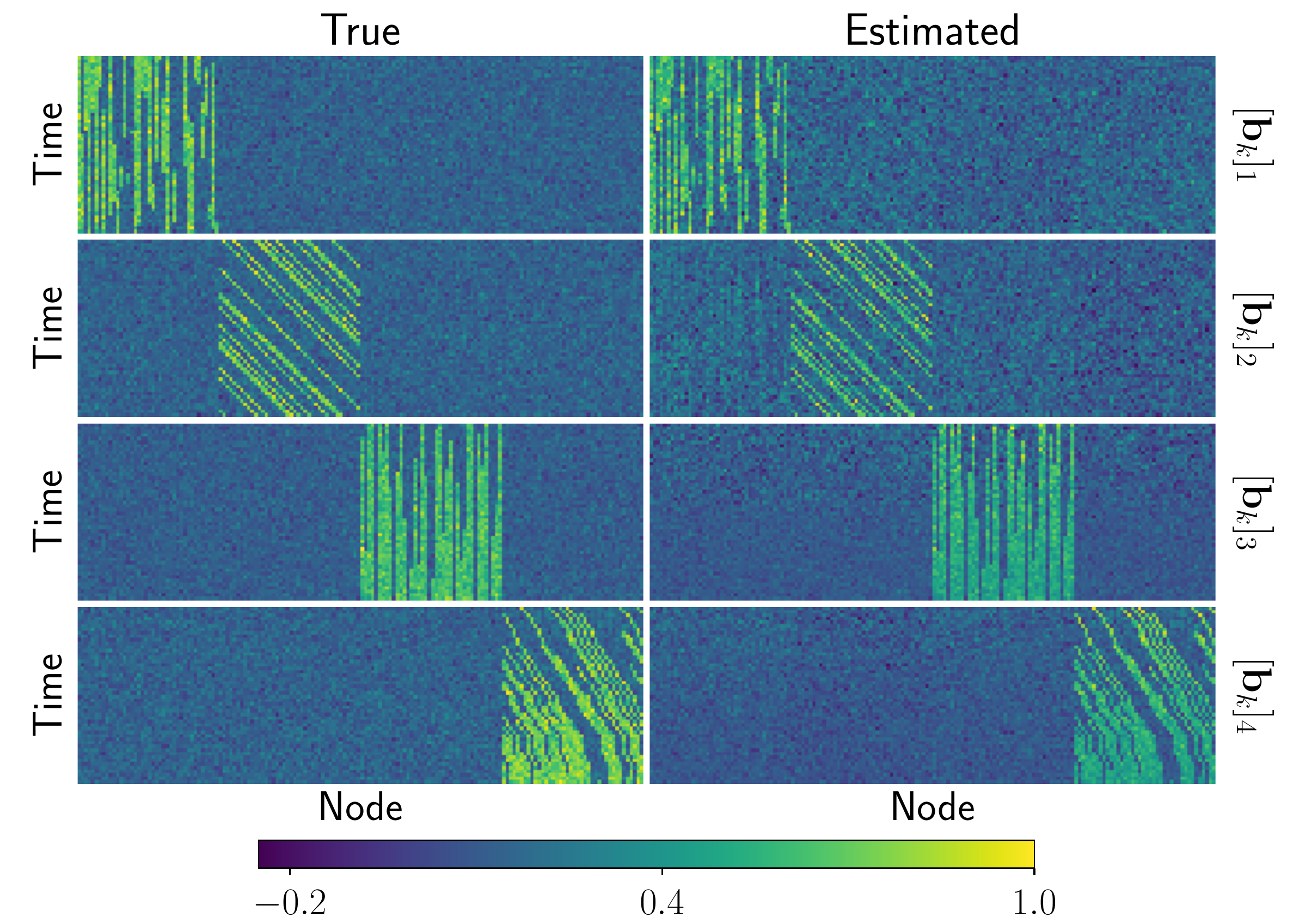}}
\end{minipage}
\caption{\ninept{True \(\Mn{B}{k}\) factor vectors (left) and the ones captured by PARAFAC2 (right) (noisy case). Factor vectors are shown as heat maps, where brighter colors represent active nodes.}}
\label{fig:B}
\end{figure}

The results also demonstrate that there is room for improvement for PARAFAC2, especially when the PARAFAC2 constraint is not satisfied, i.e. the evolving network case. Note that PARAFAC2 does not reveal the true components perfectly when $\Mn{B}{k}$s have an evolving network structure, even in the noise-free case. This challenge stems from the PARAFAC2 constraint, which requires that the 2-norm of the network factor vectors are constant in time. If a network grows or shrinks in density, the resulting change in norm cannot be represented in $\M{B}$, but rather in $\M{C}$. In the noisy case, we observe similar performance for both random and time-evolving network data. \cref{fig:AandC} and \cref{fig:B} illustrate how well PARAFAC2 performs in terms of capturing the true patterns in the noisy data.
\begin{table*}[t]
\footnotesize
\centering
\caption{\ninept{The mean performance of PARAFAC2 (PF2) and CP for different setups.}}\label{tab:results}
\begin{tabular}{@{}lccccccccccccc@{}}
\toprule
 & & & \multicolumn{2}{c}{\textbf{Fit [\%]}} & \multicolumn{2}{c}{\textbf{Clustering Acc [\%]}} &
\multicolumn{2}{c}{\textbf{FMS}\(_{\M{A}}\)} & \multicolumn{2}{c}{\textbf{FMS}\(_{\M{B}}\)} &
\multicolumn{2}{c}{\textbf{FMS}\(_{\M{C}}\)}\\

\cmidrule(lr){4-5}\cmidrule(lr){6-7}\cmidrule(lr){8-9}\cmidrule(lr){10-11}\cmidrule(lr){12-13}
\textbf{Noise} & \textbf{C setup} & \textbf{B Setup} & \multicolumn{1}{c}{CP} & \multicolumn{1}{c}{PF2} & \multicolumn{1}{c}{CP} & \multicolumn{1}{c}{PF2} & \multicolumn{1}{c}{CP} & \multicolumn{1}{c}{PF2}  & \multicolumn{1}{c}{CP} & \multicolumn{1}{c}{PF2} & \multicolumn{1}{c}{CP} & \multicolumn{1}{c}{PF2}\\

\cmidrule(r){1-1}\cmidrule(r){2-2}\cmidrule(lr){3-3}\cmidrule(lr){4-4}\cmidrule(lr){5-5}\cmidrule(lr){6-6}\cmidrule(lr){7-7}\cmidrule(lr){8-8}\cmidrule(lr){9-9}\cmidrule(lr){10-10}\cmidrule(lr){11-11}\cmidrule(lr){12-12}\cmidrule(lr){13-13}
0 & Random & Network &              75.1 &    100.0 &                          89.3 &    94.6 &       0.75 &     0.98 &       0.58 &     0.97 &       0.87 &     0.98 \\
  &        & Random &              13.2 &   100.0 &                          82.5 &    90.9 &       0.27 &     1.00 &       0.01 &     1.00 &       0.59 &     1.00 \\
  & Trends & Network &              80.7 &    100.0 &                          82.0 &    92.0 &       0.52 &     0.97 &       0.54 &     0.98 &       0.86 &     0.98 \\
  &        & Random &              15.5 &   100.0 &                          77.8 &    91.8 &       0.17 &     1.00 &       0.01 &     1.00 &       0.47 &     1.00 \vspace{0.5em}\\
 
0.33 & Random & Network &              67.8 &    91.1 &                          88.1 &    95.0 &       0.74 &     0.98 &       0.58 &     0.95 &       0.87 &     0.98 \\
     &        & Random &              11.9 &    91.1 &                          82.4 &    93.0 &       0.27 &     0.97 &       0.01 &     0.92 &       0.58 &     1.00 \\
     & Trends & Network &              72.8 &    91.1 &                          85.9 &    95.0 &       0.52 &     0.95 &       0.52 &     0.92 &       0.86 &     0.98 \\
     &        & Random &              14.0 &    91.1 &                          77.4 &    90.3 &       0.17 &     0.95 &       0.01 &     0.90 &       0.47 &     0.99 \\
\bottomrule

\end{tabular}
\end{table*}

\subsection{Real data: fMRI}
Our motivation for exploring PARAFAC2 is to understand task-related dynamic connectivity in the brain and how that differs between controls and patients suffering from a psychiatric disorder. Here, we use PARAFAC2 to analyze fMRI data from the MCIC collection \cite{fMRI:MCIC}, which contains fMRI scans from  patients with schizophrenia and healthy controls for different tasks. For this work, we used the sensory motor task (SM) data from the 3T fMRI scanners at the University of Iowa and Minnesota\footnote{We have excluded the data from other sites either due to scanner difference or observed site differences in our preliminary analysis.}. During the SM task, subjects were equipped with headphones that played sounds with increasing pitch, with a resting period between each sound. Subjects were instructed to push a button each time they heard a sound. 

To construct the data tensor, we first extracted the fractional amplitude of low-level fluctuations (fALFF) \cite{fmri:falff:ZoEtAl08} signal from sliding time windows of the blood-oxygen-level-dependent signal. The fALFF is calculated in three steps: (1) Discard high and low frequency components to remove noise and the signal from the vascular system. (2) Sum the square root of the frequency components to get the amplitude of low-frequency fluctuation. (3) Divide by the total sum of frequencies in the window. This approach provides a time-evolving measure of brain activity within each voxel. The window size and stride length were chosen so that each time window, corresponding to 16 seconds, contains precisely one block of task or rest with no overlap between the windows. The data tensor was constructed using the fALFF values for the voxels corresponding to gray matter as a feature vector for each time window and each subject. The constructed tensor is in the form of $145$ \emph{subjects} (of which, 90 of them are healthy controls) by $63652$ \emph{voxels} by $14$ \emph{time windows}.

\subsubsection{Results for fMRI data analysis}

\begin{figure}[t]
    \centering
    \includegraphics[width=1.\linewidth,trim=10 21 10 10,clip]{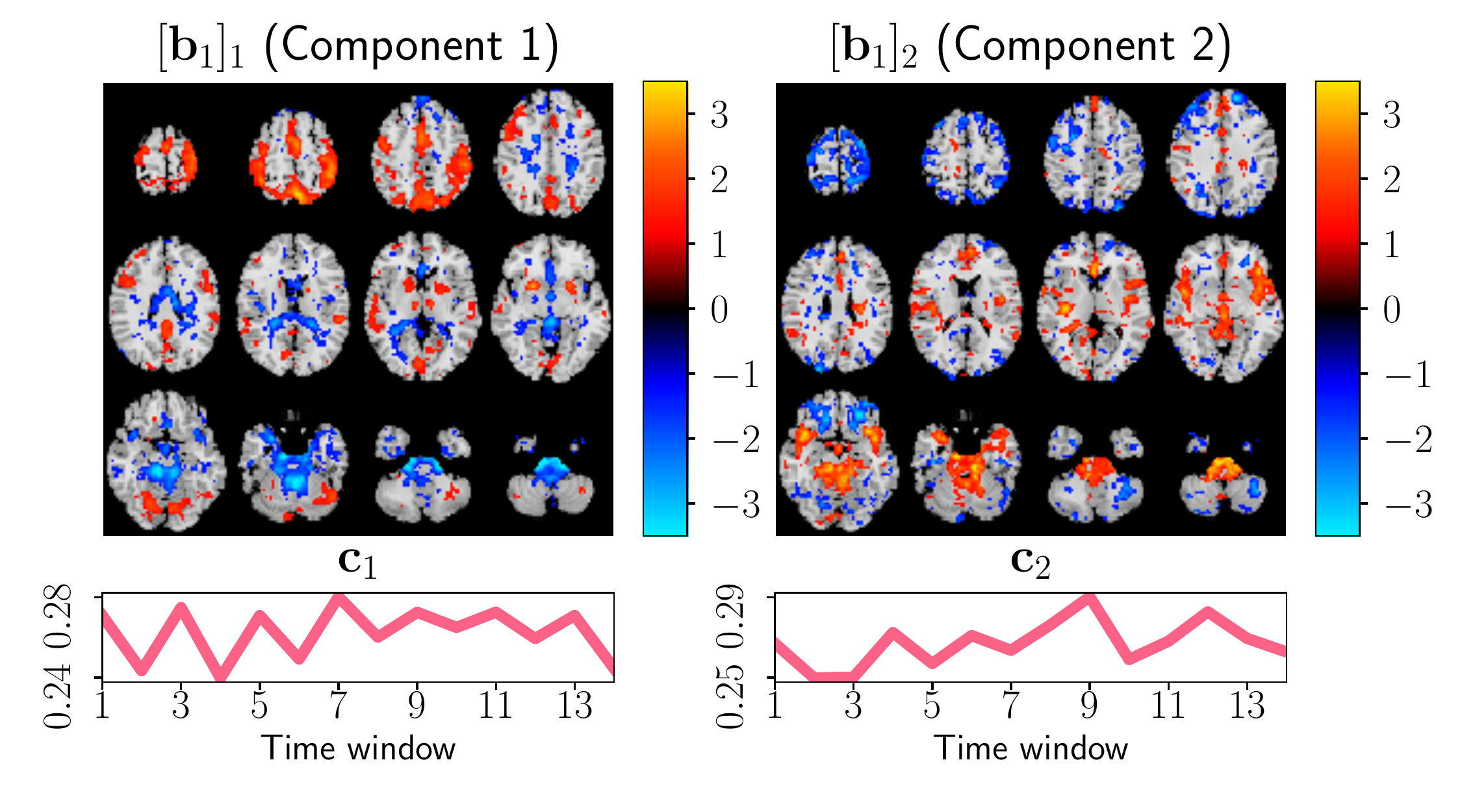}
    \caption{\ninept{Factor vectors in \emph{voxels} and \emph{time windows} modes of 2-component PARAFAC2 model. In \emph{voxels} mode, only the first time window, i.e., \(\Mn{B}{1}\), is visualized. The $p$-values are \(2.8 \times 10^{-4}\) and \(4.1 \times 10^{-2}\) for component 1 and 2, respectively.}}
    \label{fig:fmri.results.pf2}
\end{figure}

\begin{figure}[t]
    \centering
    \includegraphics[width=1.\linewidth,trim=10 21 10 10,clip]{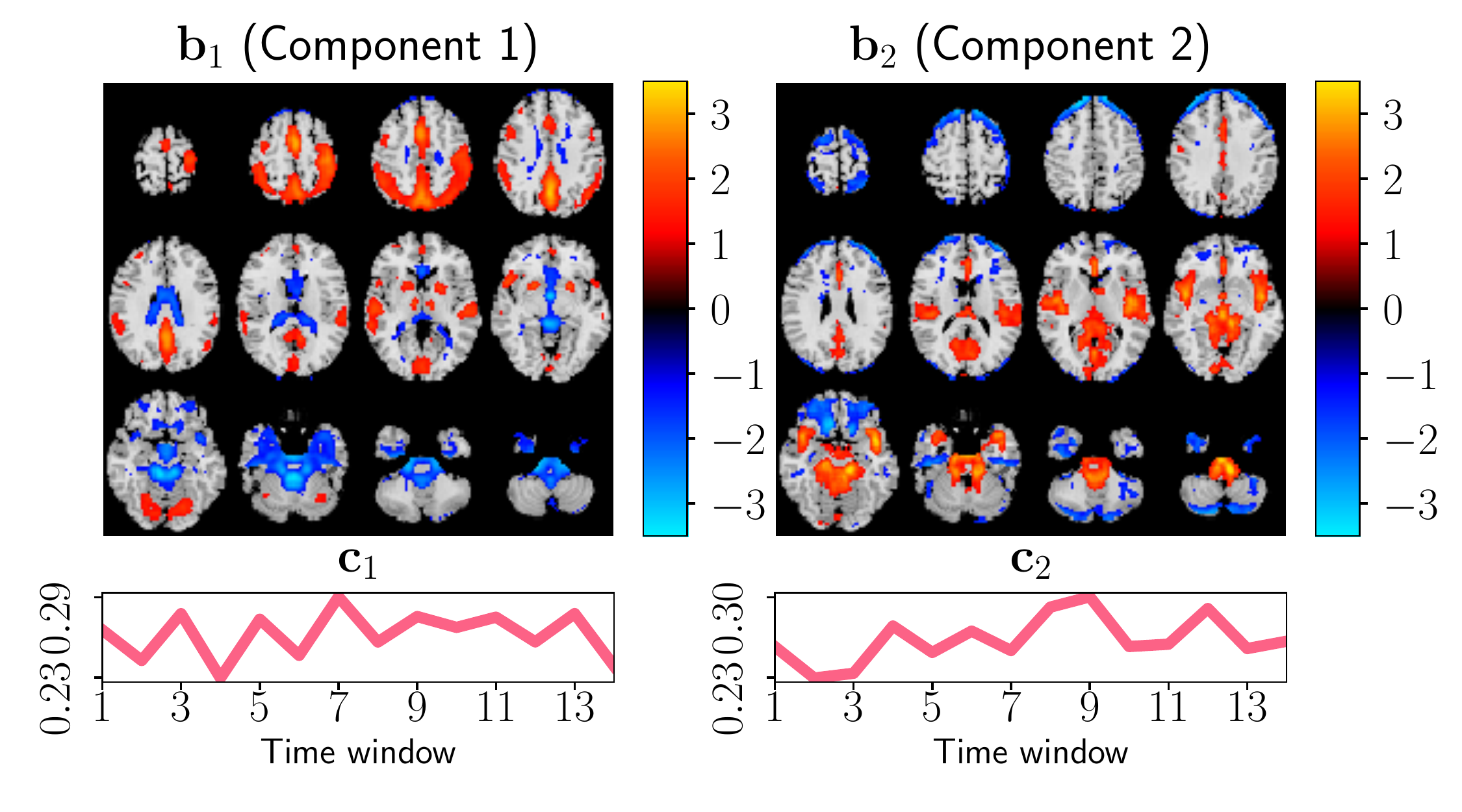}
    \caption{\ninept{Factor vectors in \emph{voxels} and \emph{time windows} mode of a 2-component CP model. The  $p$-values are \(6.2 \times 10^{-4}\) and \(3.7 \times 10^{-1}\) for component 1 and 2, respectively.}}
    \label{fig:fmri.results.cp}
\end{figure}
Before the analysis, the fMRI tensor was preprocessed by subtracting the mean fALFF signal across the \emph{voxels} mode, and dividing each \emph{voxels} mode fiber by its norm. The preprocessed tensor was then modelled using  PARAFAC2 and CP such that \(\M{A}\), \(\Mn{B}{k}\)(or \(\M{B}\) matrix for CP) and  \(\M{C}\) correspond to \emph{subjects}, \emph{voxels} and \emph{time windows}, respectively. To resolve the sign indeterminacy, we imposed non-negativity constraints on \(\M{C}\) for the PARAFAC2 model. Previously, PARAFAC2 was used for fMRI data analysis \cite{parafac2:fMRI:MaChNaMo17}, where each slice $\TFS{X}{k}$ corresponded to raw signals from a single subject and patterns in time and/or voxels were allowed to change across subjects. Instead, we construct a tensor based on features capturing the dynamic activity within each window and model the tensor using PARAFAC2, letting the voxel patterns ($\Mn{B}{k}$) change in time, rather than across subjects.

Once the models were fit, we performed a two-sample $t$-test on each column of the \emph{subjects} mode factor matrix. We achieved the lowest $p$-values using 2-component CP and 2-component PARAFAC2 models illustrated in \cref{fig:fmri.results.pf2,fig:fmri.results.cp}. Spatial maps are plotted using the patterns from the \emph{voxels} modes as z-maps and thresholding at $|z| \geq 1.2$ such that red voxels indicate an increase in controls over patients and blue voxels indicate an increase in patients over controls.

Both CP and PARAFAC2 reveal activity in the sensorimotor cortex (component 1) and auditory cortex (component 2), two brain regions known to be engaged by the SM task \cite{fMRI:MCIC}. The sensorimotor component has significantly lower activation for schizophrenic patients than  healthy controls, i.e., entries of the corresponding \emph{subjects} mode vector are positive for both groups but statistically significantly higher in controls than patients. That makes sense since it is difficult for patients to follow the task; hence the activation is lower. Temporal profiles, especially for the first component, reflect the on-off task pattern. While both models reveal similar spatial networks, PARAFAC2 captures the temporal evolution of these spatial networks (see supplementary material\footnote{\url{https://github.com/marieroald/ICASSP20}} for videos showing the temporal evolution). We also observe that due to the flexibility of PARAFAC2, the PARAFAC2 spatial maps are noisier compared to the ones captured by the CP model.

\section{Conclusion}
In this work, we have demonstrated that the tensor factorization model PARAFAC2 can recover the underlying patterns and their evolution from dynamic data, even for data not strictly following the PARAFAC2 constraints. Moreover, PARAFAC2 shows promising performance for detecting task-related brain connectivity and its evolution from fMRI data. As future work, we plan to relax the PARAFAC2 constraints and incorporate prior information about the evolving network structure (e.g., temporal and spatial smoothness) by regularizing the model to improve the recovery of underlying patterns. 

\section{Acknowledgments}
We thank Rasmus Bro for helpful discussions on PARAFAC2. We also would like to acknowledge Khondoker Hossain for his help during fMRI preprocessing.


\bibliographystyle{IEEEbib}
\bibliography{refs}

\end{document}

%% file: pre-packages.tex
\usepackage{amsmath}

\usepackage{amsfonts}

\usepackage{amssymb}

\usepackage{stmaryrd}

\usepackage[mathscr]{eucal}

\usepackage{amsbsy}

\usepackage{bm}

\usepackage{bbding}

\usepackage{array}

\usepackage{tabularx}

\usepackage{ctable}

\usepackage[neveradjust]{paralist}

\usepackage{fancyvrb}

\usepackage{caption}

\usepackage{placeins}

\usepackage{ifpdf}
\ifpdf
\DeclareGraphicsExtensions{.pdf,.png}
\else
\DeclareGraphicsExtensions{.eps}
\fi

\usepackage{algpseudocode}


%% file: pre-definitions.tex
\newcommand{\All}{\emph{:}}

\newcommand{\Tra}{^{{\sf T}}} 
\newcommand{\V}[1]{{\bm{\mathbf{\MakeLowercase{#1}}}}} 
\newcommand{\Vhat}[1]{{\bm \hat{\mathbf{\MakeLowercase{#1}}}}} 

\newcommand{\Oprod}{\circ} 

\newcommand{\M}[1]{{\bm{\mathbf{\MakeUppercase{#1}}}}} 
\newcommand{\MC}[2]{\V{#1}_{#2}} 
\newcommand{\MhatC}[2]{\Vhat{#1}_{#2}} 
\newcommand{\MR}[2]{\V{#1}_{#2 \All}} 
\newcommand{\Mn}[2]{\M{#1}^{(#2)}} 
\newcommand{\MnTra}[2]{\M{#1}^{(#2){{\sf T}}}} 

\newcommand{\T}[1]{\boldsymbol{\mathscr{\MakeUppercase{#1}}}} 
\newcommand{\That}[1]{\boldsymbol{\hat \mathscr{\MakeUppercase{#1}}}} 
\newcommand{\TFS}[2]{\M{#1}_{{#2}}} 

\newcommand{\norm}[1]{\left\lVert \, #1 \, \right\rVert}
\newcommand{\fnorm}[1]{\left\lVert \, #1 \, \right\rVert_{F}}



%% file: pre-declarations.tex
\newcommand{\email}[1]{\href{mailto:#1}{\nolinkurl{#1}}}

\newcommand{\Sec}[1]{\hyperref[sec:#1]{\S\ref*{sec:#1}}} 
\newcommand{\Section}[1]{\hyperref[sec:#1]{Section~\ref*{sec:#1}}} 
\newcommand{\AppFull}[1]{\hyperref[sec:#1]{Appendix~\ref*{sec:#1}}} 
\newcommand{\Eqn}[1]{\hyperref[eq:#1]{(\ref*{eq:#1})}} 
\newcommand{\Fig}[1]{\hyperref[fig:#1]{Figure~\ref*{fig:#1}}} 
\newcommand{\Tab}[1]{\hyperref[tab:#1]{Table~\ref*{tab:#1}}} 
\newcommand{\Thm}[1]{\hyperref[thm:#1]{Theorem~\ref*{thm:#1}}} 
\newcommand{\Cor}[1]{\hyperref[cor:#1]{Corollary~\ref*{cor:#1}}} 
\newcommand{\Alg}[1]{\hyperref[alg:#1]{Algorithm~\ref*{alg:#1}}} 
\newcommand{\Def}[1]{\hyperref[def:#1]{Definition~\ref*{def:#1}}} 

\newcommand{\Real}{{\mathbb R}}

